\documentclass[twocolumn,superscriptaddress,floatfix,prl]{revtex4}
\usepackage[latin9]{inputenc}
\usepackage{graphicx}
\makeatletter

\usepackage{amsmath}

\newcommand{\avg}[1]{\left< #1 \right>} 
\newcommand{\bea}{\begin{eqnarray}}
\newcommand{\eea}{\end{eqnarray}}
\newcommand{\beq}{\begin{equation}}
\newcommand{\eeq}{\end{equation}}
\newcommand{\benu}{\begin{enumerate}}
\newcommand{\enu}{\end{enumerate}}

\newcommand{\ket}[1]{\left| #1 \right>} 
\newcommand{\bra}[1]{\left< #1 \right|} 
\makeatother

\begin{document}


\title {Gate dependent electronic Raman scattering in graphene}

\author{E. Riccardi}
\email{elisa.riccardi@univ-paris-diderot.fr}
\affiliation{Laboratoire Mat\'eriaux et Ph\'enom\`enes Quantiques (UMR 7162 CNRS), Universit\'e Paris Diderot-Paris 7, B\^atiment Condorcet, 75205 Paris Cedex 13, France}

\author{M.-A. M\'easson}
\affiliation{Laboratoire Mat\'eriaux et Ph\'enom\`enes Quantiques (UMR 7162 CNRS), Universit\'e Paris Diderot-Paris 7, B\^atiment Condorcet, 75205 Paris Cedex 13, France}

\author{M. Cazayous}
\affiliation{Laboratoire Mat\'eriaux et Ph\'enom\`enes Quantiques (UMR 7162 CNRS), Universit\'e Paris Diderot-Paris 7, B\^atiment Condorcet, 75205 Paris Cedex 13, France}

\author{A. Sacuto}
\affiliation{Laboratoire Mat\'eriaux et Ph\'enom\`enes Quantiques (UMR 7162 CNRS), Universit\'e Paris Diderot-Paris 7, B\^atiment Condorcet, 75205 Paris Cedex 13, France}

\author{Y. Gallais}
\email{yann.gallais@univ-paris-diderot.fr}
\affiliation{Laboratoire Mat\'eriaux et Ph\'enom\`enes Quantiques (UMR 7162 CNRS), Universit\'e Paris Diderot-Paris 7, B\^atiment Condorcet, 75205 Paris Cedex 13, France}



\begin{abstract}

We report the direct observation of polarization resolved electronic Raman scattering in a gated monolayer graphene device. The evolution of the electronic Raman scattering spectra with gate voltage and its polarization dependence are in full agreement with theoretical expectations for non-resonant Raman processes involving interband electron-hole excitations across the Dirac cone. We further show that the spectral dependence of the electronic Raman scattering signal can be simply described by the dynamical polarizability of graphene in the long wavelength limit. The possibility to directly observe Dirac fermion excitations in graphene opens the way to promising Raman investigations of electronic properties of graphene and other 2D crystals.

\end{abstract}

\maketitle


 Graphene is a unique system consisting in a single layer of honeycomb carbon lattice. The exceptional physical properties of graphene are determined by its peculiar electronic structure near the Dirac point, where the linear dispersion allows to describe the graphene electrons as massless relativistic particles \cite{Novoselov2004,Novoselov2005, Zhang2005}.  Optical spectroscopies are an attractive alternative to electrical transport for probing electronic excitations and excited-state properties of graphene \cite{Specrev2012,Basov2014}, and both infrared (IR) and THz spectroscopies have been applied successfully to probe carrier dynamics near the Dirac point \cite{Li2008,Wang2008,Mak2008,Horng2011,Yan2011,Ren2012,Maeng2012}.  Some of these studies were performed on gated graphene devices, allowing a fine tuning of the Fermi level in order to study both intraband (Drude) excitations in the THz regime and interband excitations on the mi-infrared regime. Mid-infrared measurements were also extended at high magnetic field where Landau level transitions could be observed \cite{Sadowski2006,Jiang2007,Henricksen2010}. However, because of inherent limitations due to the large photon wavelength in the IR, most of these studies have been limited to relatively large area samples such as CVD grown graphene, which are still limited in terms of mobility, or graphene on SiC, where the carrier density cannot be tuned by a gate voltage. This has somewhat hampered the study of electronic interaction effects by spectroscopy as the extracted scattering rates are probably dominated by disorder effects.
\par   
Raman inelastic light scattering is in principle an attractive alternative to the above spectroscopies because it is a low frequency spectroscopy (from the mid-IR down to the THz regime) and due to the fact that it uses visible photons, it has a sub-micron spatial resolution, therefore allowing the study of a wider array of graphene devices including the cleanest ones.  
Raman spectroscopy holds indeed a privileged position in the study and characterization of graphitic materials. Up to now, its use in graphene has been almost exclusively limited to the study of optical phonons, whose properties as a function of  the number of layers, chemical doping, gate voltage and stress have been extensively investigated \cite{Malard2009,Ferrari2013}. In particular studies on gated graphene devices were able to extract information on the electronic properties of graphene both at zero \cite{Ferrari2006,Yan2007,Pisana2007} and finite \cite{Faugeras2009,Yan2010,Remi2014} magnetic field via electron-phonon coupling effects which strongly renormalize the G-band optical phonon self-energy. More recently the observation of electronic Raman scattering by inter-Landau level excitations was reported at high-magnetic fields \cite{Faugeras2011,Berciaud2014,Faugeras2015}. Despite these advances, direct observation of electronic Raman scattering (ERS) at zero-magnetic field has remained up to now rather elusive. A recent work has shown that the background of the Raman spectra of graphene is strongly dependent on the nature of the substrate, making it difficult to isolate the ERS contribution \cite{Ponosov2015}. Contrary to the case of high magnetic field where sharply defined Landau levels develop in the the electronic structures, the expected ERS spectrum at zero-field is almost featureless, making it difficult to be distinguished from the background signal. In addition, contrary to semi-conductor heterostructures with a direct optical band-gap, and also to carbon nanotubes \cite{Farhat2011}, the ERS process in graphene is, except for very high electron or hole dopings, non-resonant. It has thus remained unclear whether the ERS cross-section for a one atom thick graphene layer is large enough to be detected and extracted from the background signal inherent to any Raman experiment.
\par  
In this letter we report the unambiguous observation of ERS signal at zero magnetic field in a gated monolayer graphene device. The evolution of the ERS spectra and its polarization dependence are in full agreement with theoretical expectations of the evolution of interband electron-hole excitations upon varying gate voltage \cite{KashubaFalko2009,KashubaFalko2012}. The observed ERS continuum is weak, about 100 times weaker than the 2D optical phonon band. It displays a suppression due to Pauli blocking at a threshold close to the frequency $\omega = 2 E_F $, which shifts under the application of a gate voltage. Polarization resolved measurements indicate that the ERS signal has $A_{2g}$ symmetry as expected for vertical interband transitions across the Dirac cone in graphene \cite{KashubaFalko2009}. The extracted evolution of the Fermi energy with the gate voltage agrees very well with the estimated capacitance of the device and the Fermi velocity of graphene. 
The evolution of the ERS continuum is also entirely consistent with the observed broadening of the G-band optical phonon, providing an unified picture of the two processes which are both controlled by the electronic polarizability of graphene \cite{Ando2006}.    


The graphene-based device studied was produced by exfoliation of natural graphite. Electrical contacts were first produced using e-beam lithography and Pd deposition on an oxidized Si wafer SiO$_2$ $\sim$ 280~ nm). The pre-identified graphene flake was then positioned using a dry transfer technique \cite{Dean2010} on the top of the Si/SiO$_2$ device. The resulting structure is relatively standard and allows to apply a gate voltage between the graphene sample and the doped Si substrate that acts as a back gate. 
\par
\begin{figure}[!t]
   \centering
   \includegraphics [width=0.99\linewidth] {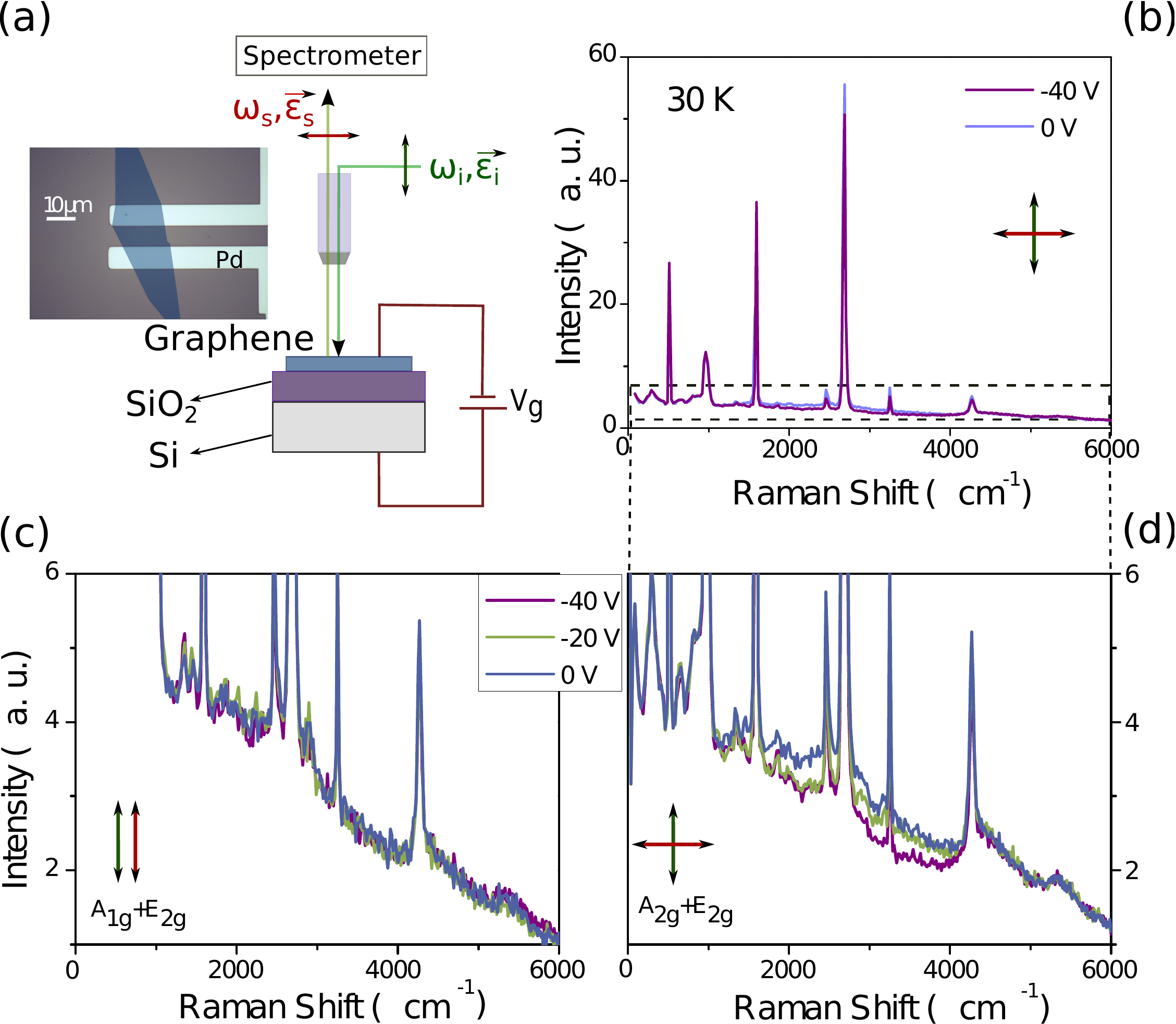}
   \caption
   {(a) Optical microscope image and schematic drawing of the device and Raman set-up. $\omega_{i,s}$ and $\varepsilon_{i,s}$ are the frequency and polarization of the in-coming and scattered photons. The Raman shift is defined as $\omega$=$\omega_i$-$\omega_s$.  (b) Cross-polarization spectra recorded at two different gate voltages: $V_g$=0 and $V_g$=-40~V in the 0 - 6000~cm$^{-1}$ frequency range. The sharp peaks are due to optical phonons (first and higher order process) of Si (below 1100~cm$^{-1}$) and of graphene (above 1100~cm$^{-1}$). (c) and (d) Polarization resolved changes in the electronic Raman continuum for parallel polarizations (c) and cross-polarizations (d) at three different gate voltages.}
   \label{fig:1}
\end{figure}

Polarization resolved Raman scattering measurements were performed using a home-built micro-Raman set-up in a back-scattering configuration equipped with a motorized $xyz$ stage with sub-micron spatial resolution. 
The $\lambda = 532~nm$ (2.33~eV) excitation line of a Diode Pumped Solid State (DPSS) Laser was focused onto the sample using a long working distance 100X objective lens. The laser spot was $\leq 1 \mu m$ and all measurements were performed with an incident laser power less than 1 mW to avoid any significant heating effects. All measurements were performed in the vacuum chamber (P $< 10^{-5}$~mbar) of a low temperature optical cryostat. The lowest cold finger temperature achieved was 30~K. The excitation beam and the collected signal were linearly polarized in order to identify the symmetry of the Raman active excitations. Using the irreducible representations of the D$_{6h}$ point group, the $A_{1g}$ and $E_{2g}$ symmetries were probed in parallel polarizations geometry and the $A_{2g}$ and $E_{2g}$ in the cross-polarizations geometry \cite{SI}. Integration times of several minutes were typically used for each spectra. The device and the optical set-up are illustrated in Fig. \ref{fig:1}(a). 



Figure \ref{fig:1}(b) shows polarization resolved spectra recorded at 0 V and -40 V, in the 0-6000~cm$^{-1}$  range and at T=30~K. The Raman signal below 1100~cm$^{-1}$ is dominated by the contribution of the doped Si layer of the substrate and does not show appreciable changes down to 80 cm~$^{-1}$ when varying the gate voltage. Above 1100~cm$^{-1}$ the easily recognizable sharp peaks are due to the first and second order optical phonon Raman processes of the graphene layer. In this work we focus on the low intensity continuum below it.   
\par
Figures \ref{fig:1}(c) and \ref{fig:1}(d) display the evolution of the polarization resolved continuum by varying the gate voltage. While the continuum is essentially independent from the gate voltage in parallel polarizations, a clear and reproducible gate-dependent effect is observed for cross-polarizations: with increasing gate voltage the continuum shows a suppression of intensity whose onset shifts to higher frequency. The suppression is not complete and concerns at most 20\% of the overall continuum intensity in cross-polarization. The gate and polarization dependences indicate two distinct contributions to the continuum intensity. One is independent of the gate voltage and dominates the spectra in parallel polarizations configuration. We assign it to residual Raman scattering signal from the Si/SiO$_2$ substrate and to luminescence coming from residual trapped impurities. Part of it could also be due to Coulomb assisted higher order ERS processes in graphene as discussed in ref. \cite{Hasdeo2013,Hasdeo2014}. The second contribution is gate dependent and only observed in cross-polarizations configuration. Its strong polarization dependence indicates that it is due to ERS by excitations of $A_{2g}$ symmetry originating from the graphene layer. As discussed below, the symmetry assignment is in agreement with the theoretical prediction of Kashuba et al. \cite{KashubaFalko2009,KashubaFalko2012} for Raman-active electron-hole interband transitions across the Dirac cone involving bands with opposite chiralities. 

\begin{figure}[!t]
   \centering
   \includegraphics [width=0.99\linewidth]{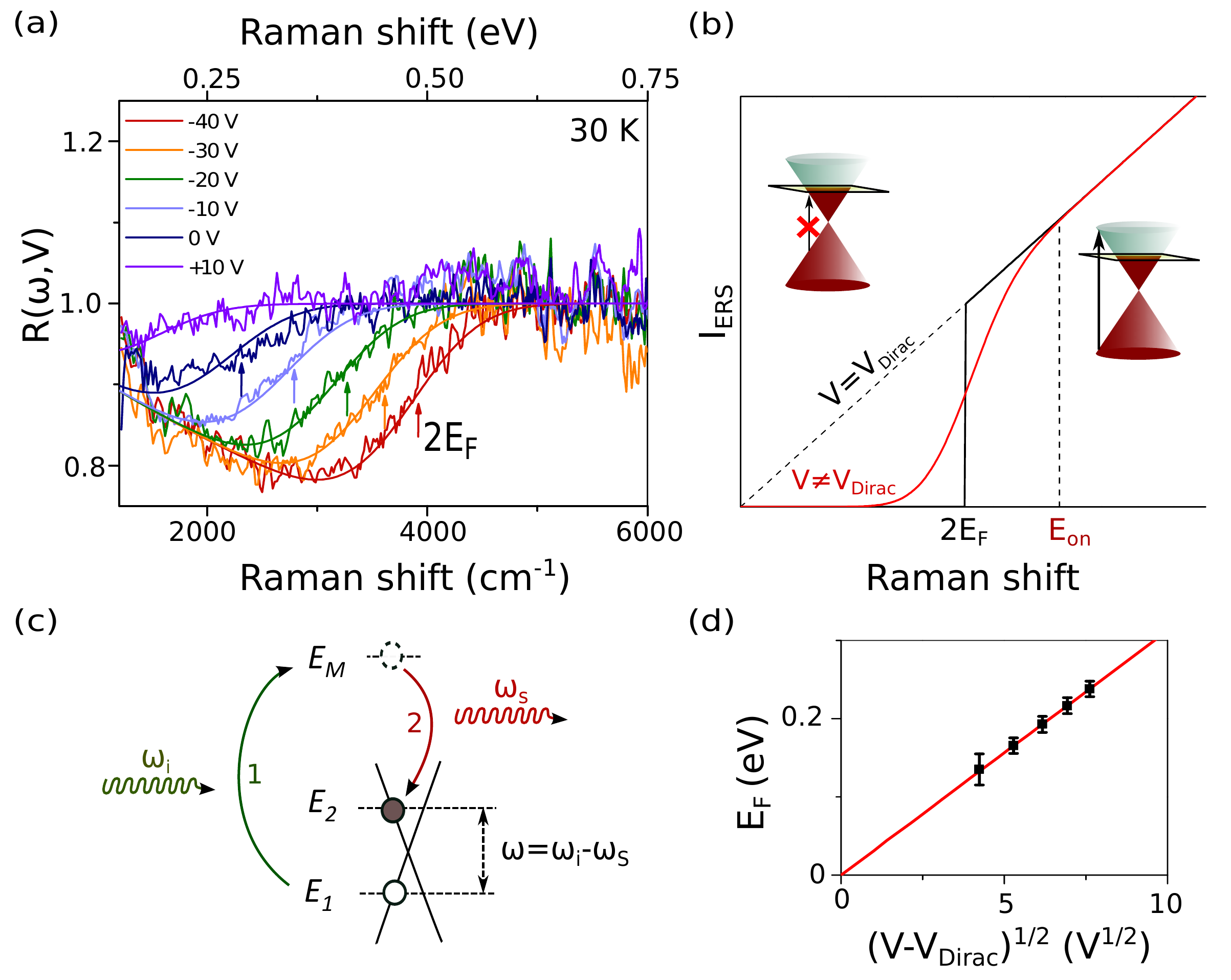}
   \caption{
   (a) Experimental and theoretical gate dependence of $R(\omega,V_g)$ at T=30~K (see text) \cite{SI3}.  (b) Evolution of the theoretical $I_{ERS}$ in graphene when the Fermi level is at the Dirac point (dotted straight line), and at finite $E_F$ (solid lines). The solid black curve is for an homogenous sample at T=0~K (Eq. \ref{eqn:fd}). The solid red line represents the finite temperature spectrum with a Gaussian distribution of Fermi Energy. The insets in (b) show schematics of vertical interband electron-hole excitations which are Pauli blocked for $\omega<2E_F$. (c) Schematic drawing of non-resonant  ($\omega_i\neq E_M-E_1$) two-step Raman process for interband electron-hole excitations via an intermediate virtual state \cite{SI}. $E_1$, $E_M$ and $E_2$ are the energies of the initial, intermediate and final electronic states. The ordering of the process is indicated explicitly: the first (second) step involves a photon induced vertical transition from the initial (virtual) state to the virtual (final) electronic state. (d) Fermi energy plotted as a function of the square root of the gate voltage. The red line is the theoretical expectation computed using the estimated gate capacitance of the device. The black dots were extracted from the $R$ data in (a) using the mid-point energy as an estimate of the inflection point \cite{SI4}.}
   \label{fig:2}
\end{figure}
\par
Proceeding with the analysis of the spectra we note that the above observations suggest the following decomposition for the continuum intensity $I$:
\begin{equation}
I(\omega,V_g)=\alpha(\omega)(I_0+I_{ERS}(\omega,V_g))
\end{equation}
where $I_{ERS}$ is the gate dependent ERS intensity from the graphene layer and $I_0$ is the gate independent intensity coming from all other sources of background as discussed above.
$\alpha(\omega)$ accounts for the instrumental spectral response which is connected to factors such as the wavelength dependence of the diffraction grating reflectivity and the CCD (charge coupled device) quantum efficiency.
It also accounts for wavelength dependent interference effects due to the presence of the substrate \cite{Yoon2009}. While all these corrections can in principle be estimated and corrected for, we choose a simpler way to extract informations on the spectral dependance of $I_{ERS}$ as a function of gate voltage. Indeed the raw spectra can be normalized with the one taken at the Dirac voltage $V_D$, I(${V_D}$), defined as the gate voltage at which the Fermi level is at the Dirac (or charge neutrality) point:  $E_F(V_g = V_D) = 0$.  We can define the ratio $R$

\begin{equation}
R(\omega,V_g)=\frac{I(\omega,V_g)}{I(\omega,V_D)}=\frac{I_0+I_{ERS}(\omega,V_g)}{I_0+I_{ERS}(\omega,V_D)}
\end{equation}

which is independent from $\alpha$ and thus free from instrumental artifacts. As pristine graphene samples are generally doped by residual impurities and/or contaminants, the Dirac voltage was estimated by following the evolution of the G-band frequency with the gate voltage \cite{Ferrari2006,Yan2007,Pisana2007}, yielding $V_D$=20~V \cite{SI2}.

For clarity, the optical phonons were first subtracted from the raw spectra, at each gate voltage, using Voigt profiles. This could however only be done reliably above~1100 cm$^{-1}$. Below 1100~cm$^{-1}$ the phonon contributions coming from the Si substrate were found to be too broad and intense to allow for an unambiguous extraction of the small gate-induced changes in the continuum underneath. The resulting phonon-free continua above 1100~cm$^{-1}$ were then divided by the spectrum at the Dirac voltage in order to obtain $R(\omega,V_g)$ which is plotted in figure \ref{fig:2}(a). As the gate voltage deviates from the Dirac voltage, $R$ is increasingly suppressed and the onset of suppression moves progressively to higher energies, reaching $\sim$ 4000~cm$^{-1}$ for $V_g$=-40~V. The behavior of $R$ bears a striking similarity with gate dependent optical conductivity data performed on similar devices \cite{Li2008,Horng2011,Ren2012}. 
\par
We now compare the experimental data with the theoretical expectations of the ERS intensity in graphene. To lowest order, the non-interacting ERS intensity arising from vertical electron-hole interband excitations in graphene reads \cite{Lu2008,Belen2009,KashubaFalko2009} \cite{SI}

\begin{multline}
I_{ERS}(\omega) = \gamma^2(\epsilon_i,\epsilon_s)\omega [f(-\frac{\hbar \omega}{2}-E_F) \\
 - f(\frac{\hbar \omega}{2}-E_F) ] \label{eqn:fd}
\end{multline}
  
where $f(E)=[1+e^{(E/k_B T)}]^{-1}$ is the Fermi-Dirac distribution. $\gamma$ is the Raman vertex which describes the electron-photon interaction process and depends on the in-coming and out-going photon polarizations. In general both direct contact processes and two-step processes involving a virtual excitation can contribute to ERS intensity \cite{SI,Wolff66}. As shown by Kashuba et al. \cite{KashubaFalko2009}, in the case of graphene and for excitation photon energies in the visible range, non-resonant two-step processes are the dominant ones for interband electron-hole excitations. This is in contrast with conventional two-dimensional electron gas in semiconductor heterostructures, where ERS is generally studied in the resonant regime, with incident photon energies $\omega_i$ tuned close to the fundamental gap of the semiconductor \cite{Pinczuk-Abstreiter}.  In graphene the associated non-resonant Raman vertex has $A_{2g}$ symmetry \cite{KashubaFalko2009}: it is non-zero only for cross linear photon polarizations in agreement with our experimental data. As shown in Fig. \ref{fig:2}(b), at the Dirac point the theoretical ERS intensity has a linear in frequency spectral dependence, while away from the Dirac point it displays a threshold at 2$E_F$ due to Pauli-blocking. Except for the linear in frequency term, the ERS frequency dependence is very similar to the optical conductivity and the approximate relation $I_{ERS}(\omega) \sim \omega\sigma_1(\omega)$ holds for graphene \cite{Belen2009}. As typical graphene samples display inhomogeneous carrier doping, spatial fluctuations of the Fermi energy were also considered and assumed to follow a Gaussian distribution function (see Fig. \ref{fig:2}(b)).
\par
In order to compare with the experimental data, a theoretical $R$ was calculated by dividing each theoretical spectra by the one at the Dirac voltage. The Fermi energies were chosen assuming the standard relationship between the gate voltage and the density:  $n=C (V_g-V_{Dirac})/e$ with $E_F = -sgn(n) \hbar v_F \sqrt{(\pi \mid n \mid)}$, using the calculated geometrical capacitance of the device ($C=110~aF$/$\mu m^2$) and a Fermi velocity $v_F$=10$^6$~ms$^{-1}$. The theoretical results are superimposed on the experimental data in Fig. \ref{fig:2}(a) for several gate voltages. The agreement between theory and experiments is remarkable given that the only free parameter is the standard deviation of the Gaussian distribution of Fermi energy. In our case $\delta E_F\sim$ 50~meV gave the best fits to the data. This value is consistent with previous estimations for supported graphene samples \cite{Martin2008,Yan2007,Berciaud2015}. We note however that others parameters, such as the finite electron lifetimes, can also contribute to the observed broadening of the 2$E_F$ threshold. We thus consider this value of $\delta E_F$ as an upper limit \cite{SI}. In Fig. \ref{fig:2}(d) we show that the Fermi energy can also be reliably obtained directly from the $R$ data by identifying the inflection point of the spectra (shown with an arrow in Fig. \ref{fig:2}(a)) with 2$E_F$.


\begin{figure}[!t]  
   \centering
   \includegraphics [width=\columnwidth]{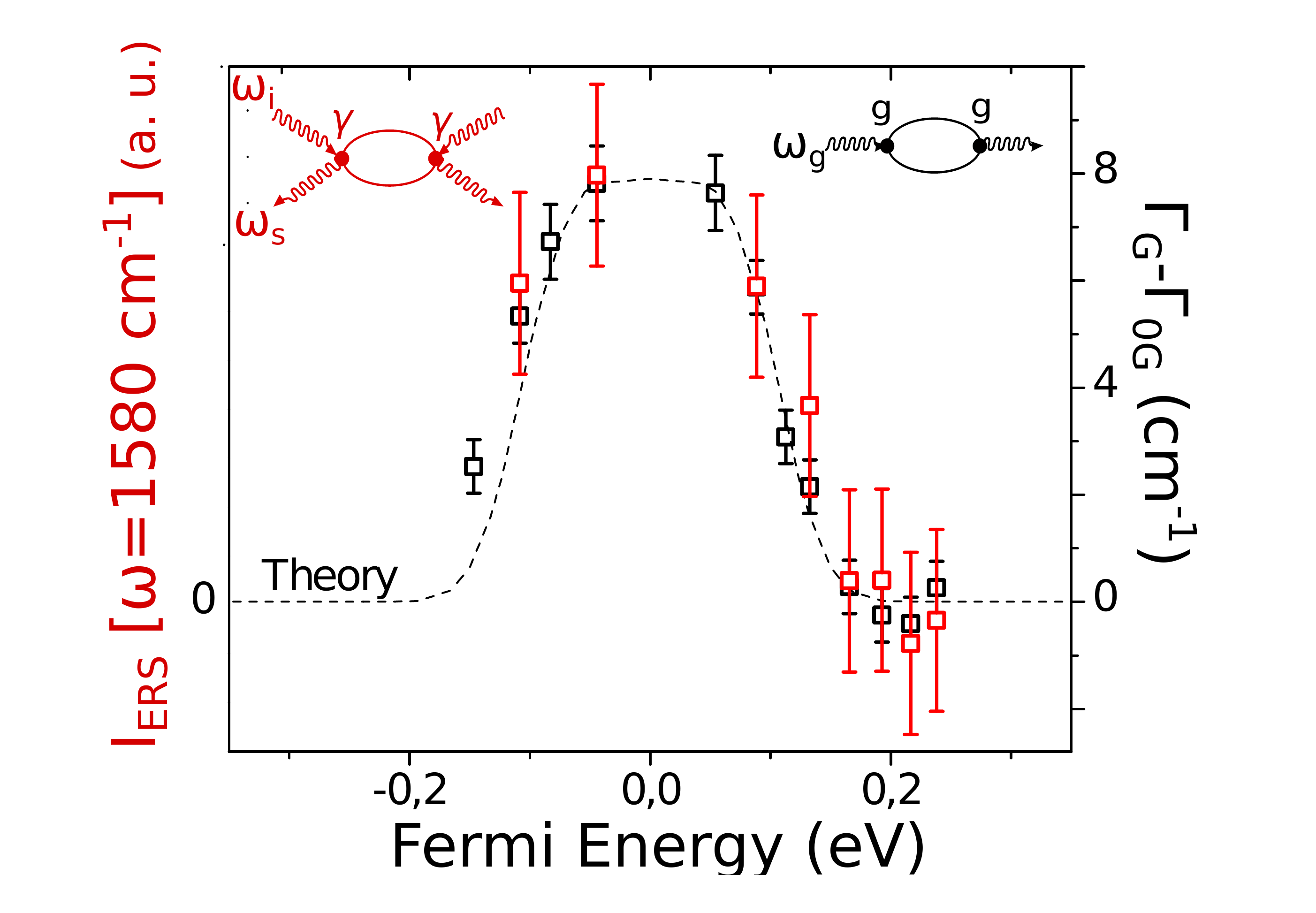}
   \caption
   {Evolution of the ERS intensity at 1580~cm$^{-1}$ (red dots) and of the G band change in linewidth $\Delta \Gamma_G$=$\Gamma_G$-$\Gamma_0$ (black dots) as a function of the Fermi energy. $\Gamma_G$ is the observed linewidth and $\Gamma_0$ a gate voltage independent contribution to the linewidth arising from lattice anharmonicity and disorder effects. The dotted line is the theoretical expectation for $I_{ERS}(1580~cm^{-1})$ after a convolution with a Gaussian distribution of Fermi energy with $\delta E_F \sim$ 50~ meV. The insets show Feynman diagrams for the ERS intensity (red) and electron-phonon induced G-band renormalization (black). See \cite{SI} for details about the ERS vertex.}
   \label{fig:3}
\end{figure}

\par
It is illuminating to draw a parallel between the observed gate dependent ERS continuum and the well-known behavior of the G-band linewidth. Indeed, apart from the Raman vertex pre-factor, the frequency dependence of the ERS intensity is essentially given by the imaginary part of the electron-polarizability of graphene, $\Pi''(\omega,q=0)$ (see e.g. \cite{Wunsch2006,Das-Sarma2007,SI}). On the other hand, as first noted by Ando \cite{Ando2006}, and observed experimentally \cite{Ferrari2006,Yan2007,Pisana2007}, due to electron-phonon coupling the G-band linewidth $\Gamma_G$ contains a contribution, $\Delta\Gamma_G$, which arises from Landau damping by electron-hole excitations processes and is given by the imaginary part of the electron-polarizability at the phonon frequency: $\Delta\Gamma_G \sim \Pi''(\omega=\omega_G,q=0)$. The only difference between ERS and G-band renormalization processes is the vertices involved: the Raman vertex $\gamma$ and the electron-phonon coupling constant $g$ respectively (see inset in Fig. \ref{fig:3}). The gate dependence of the G-band linewidth is thus directly proportional to the intensity of the ERS continuum at the phonon frequency:

\begin{equation}
\Delta\Gamma_G \sim I_{ERS}(\omega=\omega_G)
\end{equation} 

Figure \ref{fig:3} displays the gate voltage dependence of the ERS continuum taken at the G band frequency $I_{ERS}(\omega=1580~cm^{-1})$ and the change in linewidth of the G-band $\Delta \Gamma_G$, measured on the same spot but with a higher resolution. The overall agreement between both quantities provides a direct evidence of their common link to the electron-polarizability of graphene and gives an unified picture of both effects.

In conclusion we have observed gate dependent ERS signal from a single-layer graphene device. The gate voltage and light polarizations dependences of the signal are fully consistent with interband electron-hole excitations created by a non-resonant Raman process. While Raman scattering in carbon-based materials has been traditionally confined to the study of optical phonons, our work demonstrates the ability of Raman scattering in exploring electronic excitations of 2D materials even far away from resonance. It paves the way for promising future studies of interaction induced effects in cleaner devices, which have remained hitherto inaccessible to most spectroscopies. 

\acknowledgments

This work was supported by a C'Nano grant (Optographene) from Ile-de-France region and by a LABEX SEAM (Science and Engineering for Advanced Materials and devices)  grant (ANR 11 LABX 086, ANR 11 IDEX 05 02). The authors thank Indranil Paul and Dimitri Maslov for fruitful discussions, and the staff of the MPQ laboratory clean room, Stephan Suffit, Christophe Manquest and Pascal Filloux for their precious technical support.



\clearpage

\textbf{\large{Supplemental Material}}

\bigskip

In this supplemental material we provide the theoretical background on Raman selection rules and on the calculations of the electronic Raman scattering (ERS) intensity of graphene. In particular we explore finite temperature and finite electron lifetime effects. We also provide additional data on the renormalization of the phonon G-band as a function of gate voltage on the device of the main text. We also present ERS data on a second device at room temperature.

\section{Raman selection rules for D$_{6h}$ point group}
With in plane incoming and out-going photon polarizations one can access three distinct irreducible representations of the D$_{6h}$ point group: A$_{1g}$, A$_{2g}$ and E$_{2g}$. Their associated second ranked Raman tensors or vertex are given in \cite{Loudon}. They read (ignoring the out-of plane components):

\begin{center}

$\hat{\gamma}^{A_{1g}}$= $\left( \begin{array}{ccc}
a & 0  \\
0 & a 
\end{array} \right)$

\vspace{3 mm}

 $\hat{\gamma}^{A_{2g}}$= $\left( \begin{array}{ccc}
0 & b  \\
-b & 0 
\end{array} \right)$

\vspace{3 mm}

$\hat{\gamma}^{E_{2g}}$=$ \left( \begin{array}{lll}
c & 0  \\
0 & -c 
\end{array} \right)$, $\left( \begin{array}{lll}
0 & c  \\
c & 0 
\end{array} \right)$

\end{center}

\vspace{3mm}

Note that while the coefficients depend on the microscopics of light-matter interaction and therefore the type of excitations involved, phonons or electrons for e.g., the general forms of these tensors are robust because they are dictated by symmetry. The knowledge of these tensors allows to compute the Raman intensity associated to each representation $\mu$ for a given set of out going and incoming photon polarizations $\bf{\epsilon_s}$, $\bf{\epsilon_i}$:
 
\begin{equation}
I_{\mu}\propto \mid\gamma^{\mu}\mid^2=\mid{\bf{\epsilon_i}}\hat{\gamma}^{\mu}{\bf{\epsilon_s}}\mid^2
\end{equation}

Using the above equation, one can easily show that for parallel polarizations: 
\begin{equation}
I_{total}({\bf{\epsilon}}_s \parallel {\bf{\epsilon}}_i)=I_{A_{1g}}+I_{E_{2g}}
\end{equation}
while for cross polarizations:
\begin{equation}
I_{total}({\bf{\epsilon}}_s \perp {\bf{\epsilon}}_i)=I_{A_{2g}}+I_{E_{2g}}
\end{equation}

\section{Electronic Raman processes}
The non-resonant ERS intensity is directly proportional to the imaginary part of the electronic Raman response $\chi''$, which for weakly correlated systems can be written in terms of an effective density-density correlation function \cite{Wolff66,Dev-Hackl}.
\begin{equation}
I_{ERS}(\omega,q)\propto (1+n(\omega,T)) \chi''(\omega,q) 
\end{equation}
 $\mathbf{q}$ is the transferred wave-vector and $n$ the Bose-Einstein distribution whose effect is vanishingly small for $\omega >> k_BT$ and will be neglected hereafter. The Raman response function $\chi(\omega)= \avg{\rho_\mathbf{q}^{\mu}(\omega)\rho_\mathbf{q}^{\mu}(-\omega)}$ measures an effective density $\rho^{\mu}$ which is given by

\begin{equation}
\rho_{\mathbf{q}}^{\mu}=\sum_{\mathbf{k}}\gamma_{\mathbf{k,q}}^{\mu}c^+_{\mathbf{k+q}}c_{\mathbf{k}}
\end{equation}

 $\gamma_{\mathbf{k,q}}^{\mu}$ is the Raman scattering amplitude or Raman vertex whose symmetry $\mu$ is determined by the polarizations $\mathbf{\epsilon}_{i}$ and $\mathbf{\epsilon}_{s}$ of the incoming and out-going photons as discussed above. The Raman vertex $\gamma$ contains information on the two-photon interaction responsible for the Raman processes. For ERS these processes arise from (i) direct contact interaction processes and (ii) two-step processes involving virtual or real intermediate states. We note that if we replace the Raman vertex by a constant, the Raman response function is simply given by the dynamical polarizability $\Pi(\omega,\mathbf{q})$: the standard Lindhard function for non-interacting electrons \cite{Ashcroft}. 
\par
The interaction of matter with light occurs via coupling of the radiation field represented by the vector potential {\textbf{A}} to an electron with momentum ${\bf{P}}$. This interaction can be computed using the Peierls substitution ${\bf{P}}: {\bf{P}} - \frac{e{\bf{A}}}{c}$ and gives two contributions to the light matter interaction Hamiltonian:
\begin{equation}
H_A=-\frac{e}{mc}{\bf{P}}.{\bf{A}}
\end{equation}
\begin{equation}
H_{AA}= \frac{e^2}{2mc^2}{\bf{A}}^2
\end{equation}  

\begin{figure}
\centering
\includegraphics [width=0.4\textwidth] {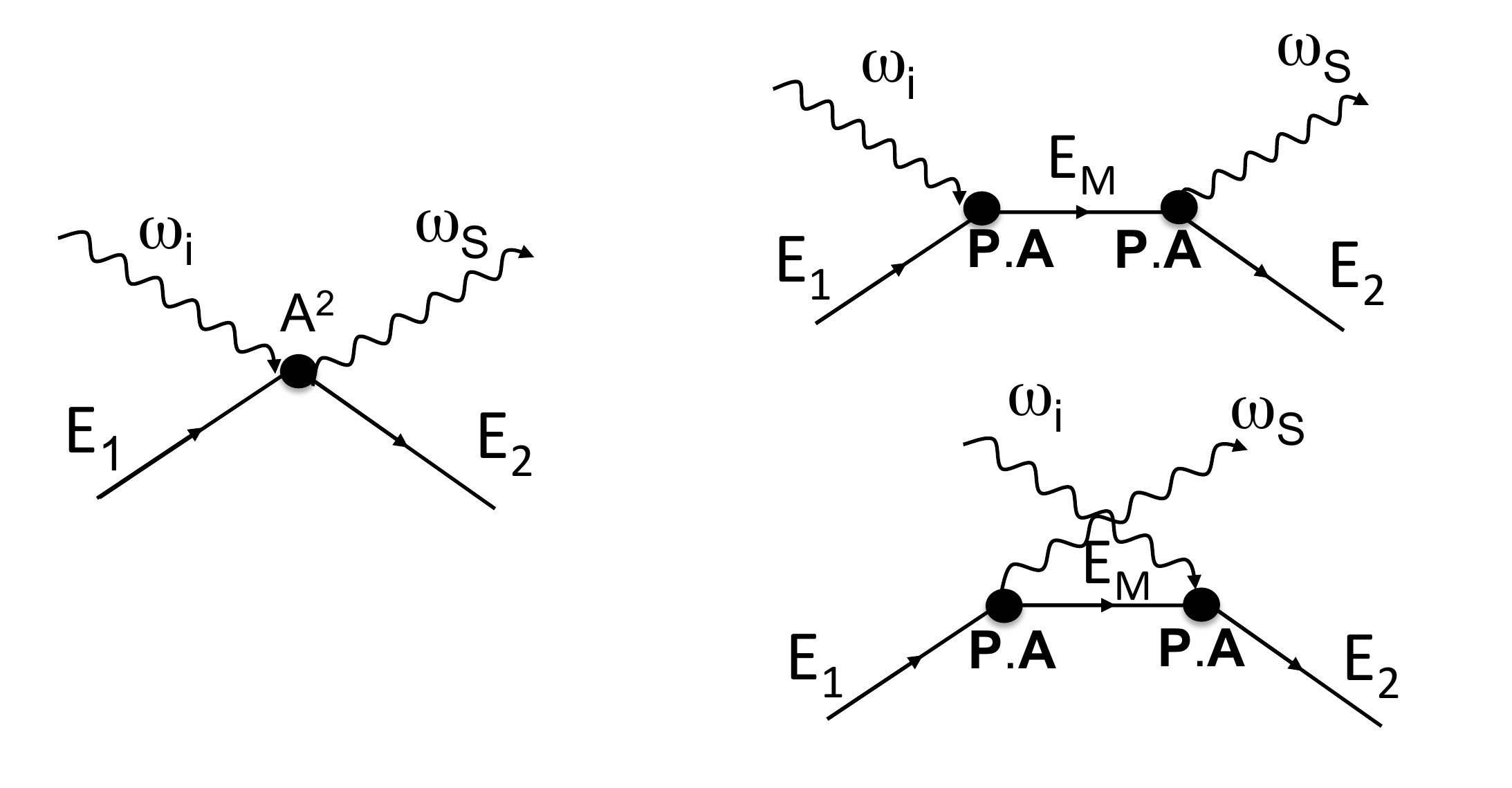}
\caption{Feynman diagrams describing one step or contact, (left) and two-step ERS processes (right). The wavy lines designate photons and the straight lines electrons.}
\label{ERS-process} 
\end{figure}

The electronic Raman cross-section corresponds to two photon processes. Its vertex can be computed by treating $H_{AA}$ to first order perturbation theory and $H_A$ to second order perturbation theory. Noting ${\bf{\epsilon}}_s$ and ${\bf{\epsilon}}_i$ the unit vectors of the in-going and out-going vector potentials ${\bf{A}}$, the general formula of the Raman vertex for non-interacting Bloch electrons can be written as (see for e.g. \cite{Wolff66}):
\begin{multline}
\gamma({\bf{\epsilon}_i},{\bf{\epsilon}}_s)\propto {\bf{\epsilon}}_i.{\bf{\epsilon}}_s+\frac{1}{m}\sum_M \frac{\bra{F}{\bf{P}}.{\bf{\epsilon}_s}\ket{M}\bra{M}{\bf{P}}.{\bf{\epsilon}_i}\ket{I}}{E_M-E_1-\hbar\omega_i} \\
+\frac{\bra{2}{\bf{P}}.{\bf{\epsilon}_i}\ket{M}\bra{M}{\bf{P}}.{\bf{\epsilon}_s}\ket{1}}{E_M-E_1+\hbar\omega_s}
\label{vertex}
\end{multline}
where $\ket{1}$, $\ket{M}$ and $\ket{2}$ are the initial, intermediate and final Bloch electron states with energies $E_1$, $E_M$ and $E_2$ respectively. These represents the lowest order ERS processes. Higher order processes involving Coulomb interaction matrix element can also contribute. However they will typically give raise to a featureless continuum with a very weak dependence on the gate voltage \cite{Hasdeo}. We therefore do not consider them here. To lowest order the Raman vertex \ref{vertex} contains two types of contributions depicted in the Feynman diagrams shown in figure \ref{ERS-process}:

\begin{enumerate}
\item the first term is a direct contact term in $A^2$ where the inelastic process occurs in a single step of light matter interaction (two photon scattering). This term is never resonant.
\item the last two terms in {\textbf{P}}.{\textbf{A}} corresponds to two-step processes: a photon emission (absorption) followed by a photon absorption (emission). Note that the intermediate electronic state $\ket{M}$ does not have to satisfy energy conservation and only wave-vector should be conserved (i.e. transitions are almost vertical). The sum is therefore over all the electronic states of the crystals which satisfy momentum conservation. The intermediate state can either virtual, (non-resonant process), or real, (resonant process). In the resonant case the incident photon energy matches an interband transition: $E_M-E_1=\hbar\omega_i$ and the vertex diverges. When finite lifetime effects are included the divergence will be replaced by a strong enhancement. This enhancement is observed for example in semiconductor heterostructures when the incoming photon energy is tuned close to the fundamental gap energy of the semiconductor host. In the case of graphene, Raman processes involving interband electron-hole excitations across the Dirac cone can only be resonant if the Fermi energy $E_F$ is on the order of the incoming photon energy $\omega_i$. In typical Raman experiment the incoming photon energy is in the near-infrared - visible range (1.5 - 3eV) so that $\omega_i>> E_F$ and resonant interband electron-hole excitation processes cannot occur. This is because both wave-vector conservation and the resonance condition on the intermediate state energy cannot be met simultaneously. This leaves non-resonant terms involving either contact interaction or two-step processes involving virtual intermediate states as the only possible processes. We note that higher order ERS processes do not have this restriction, and will be in general resonant \cite{Hasdeo}. They may contribute to the gate and symmetry independent continuum observed in our experiments.
\end{enumerate}

 \par
For non-resonant Raman scattering it is customary to compute the Raman vertex using the effective mass approximation. In the case of graphene however, the effective mass approximation gives vanishing Raman amplitude because of the linear electronic dispersion. Instead the contact and two-step processes terms have to be calculated explicitly. This calculation was performed by Kashuba et al. \cite{Kashuba} and, for interband electron-hole excitations they found a k-independent Raman vertex dominated by non-resonant two-step processes. Because of the chirality of the electronic bands of graphene, its leading term has $A_{2g}$ symmetry and therefore breaks time-reversal symmetry. In addition it scales as the square of the incoming photon energy: $\gamma^{A_{2g}}_k=\gamma_0\propto \frac{1}{\omega_L^2}$. 
\begin{figure}
 \includegraphics [width=0.4\textwidth] {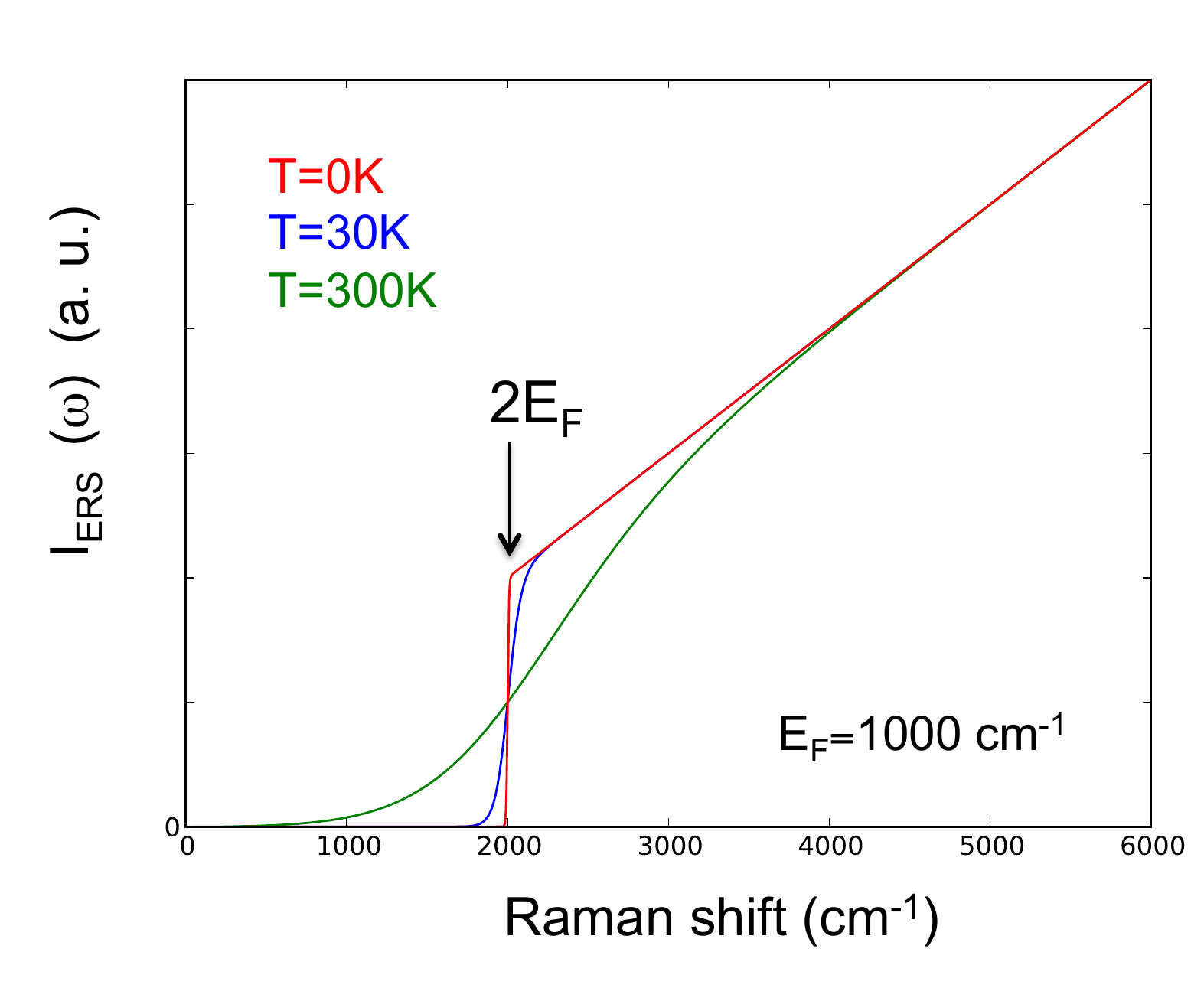}
 \caption{Theoretical ERS intensity $I_{ERS}$ from interband electron-hole excitations at different temperatures and for $E_F$=1000~cm$^{-1}$. A clean (infinite electron lifetime) and homogeneous graphene layer is assumed.}
 \label{fig1SI}
\end{figure}
\begin{figure}
 \includegraphics [width=0.4\textwidth] {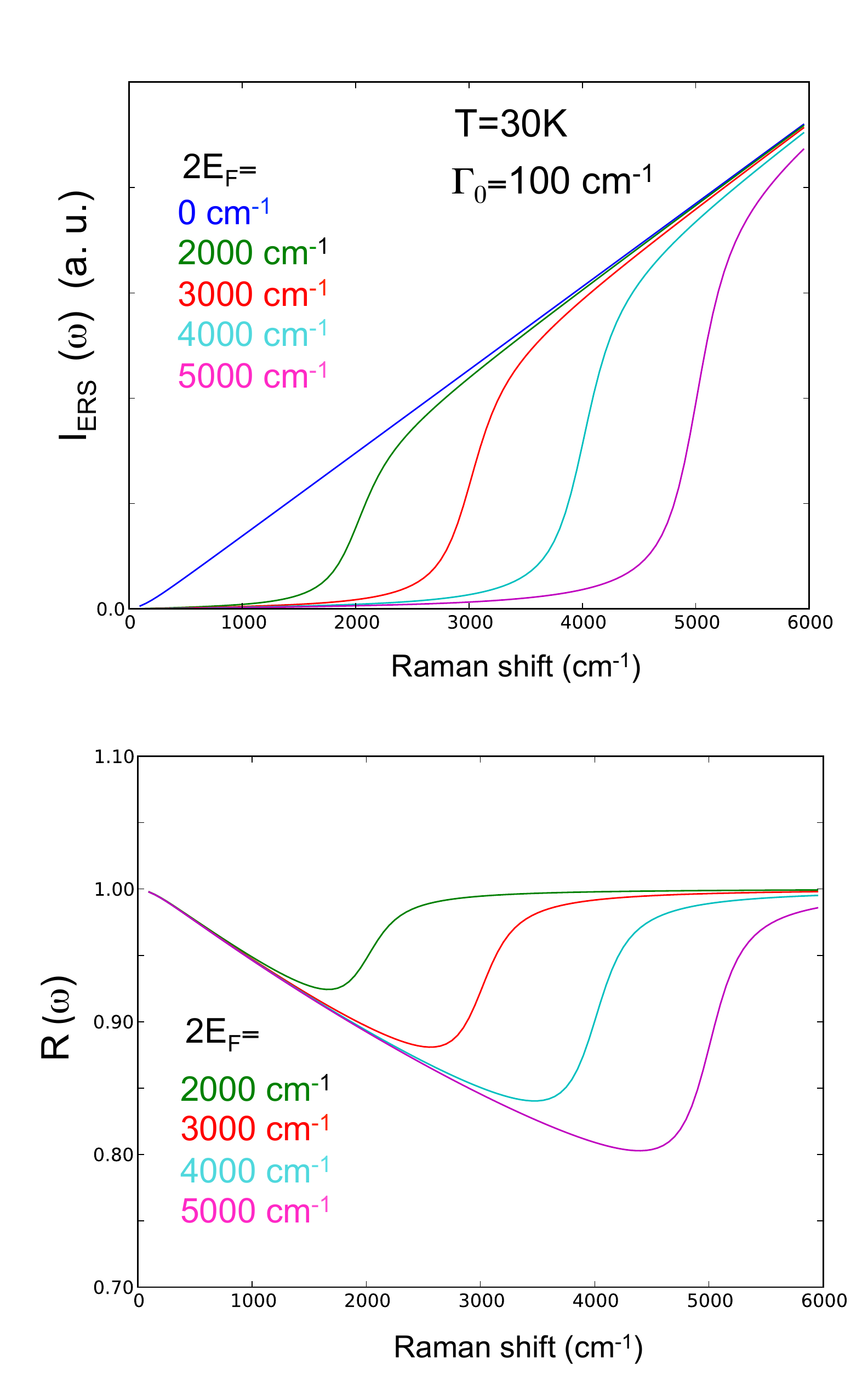}
 \caption{Theoretical ERS intensity $I_{ERS}$ and ratio $R$ from interband electron-hole excitations at 30~K for different Fermi energies. An electron scattering rate of 100~cm$^{-1}$ was assumed.}
 \label{fig2SI}
\end{figure}
\section{Electronic Raman reponse calculations}
With the knowledge of the Raman vertex $\gamma_k$, a general expression for the imaginary part of the electronic Raman response, $\chi''$, can be given via the use of the electronic spectral function $A(k,\omega)$ \cite{Dev-Hackl}

\begin{multline}
\chi''(\omega,\mathbf{q})= \sum_{s,s'} \int d\mathbf{k} \gamma^2_{\mathbf{k,q}}  \int d\epsilon [f(\epsilon)-f(\epsilon+\hbar\omega)] \\
A_s(\mathbf{k},\epsilon)A_{s'}(\mathbf{k+q},\epsilon+\hbar\omega)
\label{chi_ERS}
\end{multline}

where $s,s'=\pm 1$ refer to the two linearly dispersing bands $E_k=sv_F \lvert \mathbf{k}\rvert$, $f$ is the Fermi-Dirac distribution function $f(\epsilon)=[1+e^{\frac{\epsilon-E_F}{k_BT}}]^{-1}$ and $A_s(k,\epsilon)$ is the one-particle electronic spectral function which can be written in terms of the electron self-energy $\Sigma$:
\begin{equation}
A_s(\mathbf{k},\omega)=\frac{-2\Sigma''}{[\hbar\omega-sv_F\lvert k\rvert-\Sigma']^2+\Sigma''^2}
\end{equation}
Both intraband, s=s', and interband, s$\neq$s', can contribute to the electronic Raman response. 

\subsection{Clean graphene at finite temperature}
If we ignore the electron self-energy $\Sigma$=0 the spectral functions can be  replaced by delta functions: $A_s(\mathbf{k},\omega)=\delta(\hbar\omega-sv_F\lvert k\rvert)$. In that case the intraband term is limited to frequencies below $\mathbf{q}.\mathbf{v_F}$ \cite{Platzman}.  We note that for back-scattering configuration with a photon wave-vector perpendicular to the plane of graphene, the transferred wave vector ${\bf{q}}$ is vanishingly small and there is no contribution from intraband processes. The interband contribution however remains finite and can be calculated analytically using Eq. \ref{chi_ERS}. For $q$=0 and assuming a k-independent Raman vertex $\gamma_0$ , we have

\begin{multline}
\chi''(\omega,q=0)=\gamma_0^2\sum_{s\neq s'} \int (\frac{kdk}{2\pi})^2 \int d\epsilon [f(\epsilon)-f(\epsilon+\hbar\omega)] \\
\delta(\epsilon-sv_F\lvert k\rvert))\delta(\hbar\omega+\epsilon-s'v_F\lvert k\rvert))
\end{multline}
Integration over $k$ and $\epsilon$ gives: 

\begin{equation}
\chi''(\omega)=\gamma_0^2\omega[f(-\frac{\hbar\omega}{2})-f(\frac{\hbar\omega}{2})]
\end{equation}
which is equivalent to Eq. 3 of the main text. This expression can be compared with the one at T=0~K of Kashuba et al. \cite{Kashuba} where the Fermi-Dirac distribution functions are replaced by a step function $\theta$
\begin{equation}
\chi''(\omega)=\gamma_0^2\omega \theta(\hbar\omega-2E_F)
\end{equation}
The corresponding ERS spectrum is plotted for several temperatures for $E_F$=1000~cm$^{-1}$ in Fig. \ref{fig1SI}. We notice that for T=30~K the effect of temperature is barely noticeable with respect to T=0~K calculations. Temperature effects alone thus cannot explain the broadening of the 2$E_F$ suppression seen experimentally at 30~K. They are however more significant at 300~K.

\begin{figure*}
 \includegraphics [width=0.9\textwidth] {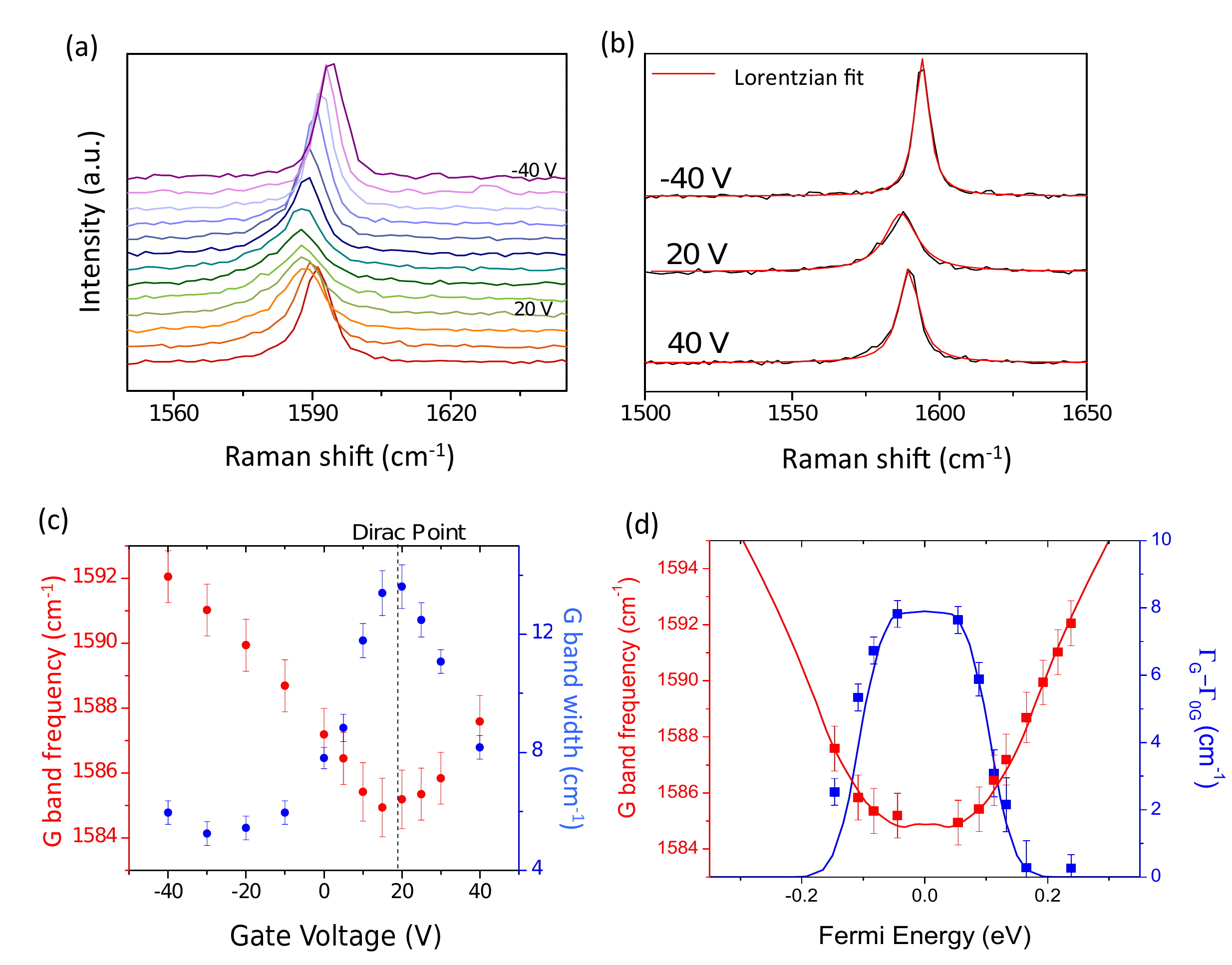}
 \caption{(a) G band phonon Raman spectra recorded at 30~K for different gate voltage values. (b) Lorentzian fits of the G band spectra at -40~V, 20~V (Dirac voltage) and 40~V. (c) Gate dependence of the G band energy and line-width, extracted from (a). The dotted line indicate the approximate position of the Dirac voltage. (d) Evolution of the G band energy and line-with with the Fermi energy extracted from the estimated capacitance and consistent with ERS data (see main text). The lines correspond to theoretical fits using the same gaussian distribution of Fermi energy as in ERS data ($\delta E_F$=50~meV) and an electron-phonon coupling strength $D$=13~eV/$\AA$ \cite{Ando,Jun, Pisana}.}
 \label{fig3SI}
\end{figure*}
\section{Finite lifetime effects}
We now take into account electron lifetimes effects more explicitly by considering the full expression for the electron spectral function including self-energy effects. In particular we explore the effect of disorder which will result in a finite frequency independent scattering rate $\Gamma_0$ which is given by the imaginary part of the self-energy $\Sigma$: $\Sigma''=\Gamma_0$. We note that electron-electron interaction can also bring additional frequency dependent contributions to the self-energy. We do not consider them here but they might important for cleaner graphene samples like suspended ones, where the impact of disorder is reduced.
\par   
In order to proceed we need an estimate of the electron scattering rate. Within a Drude model, the typical mobilities of supported graphene sample suggest a impurity induced scattering rate $\Gamma_0\sim$~100 cm$^{-1}$ \cite{Basov-review}. This value is confirmed by THz measurements of the Drude response in CVD graphene samples \cite{Horng}. This value was therefore used and inserted in the electron spectral functions which are now Lorentzian with a line-width given by $\Gamma_0$. The corresponding ERS intensity and ratio $R$ were numerically computed and are shown for different values of the Fermi energy $E_F$ in Fig. \ref{fig2SI}. The broadening of the 2$E_F$ suppression is still insufficient to account for the data hinting that inhomogeneous broadening due to non-uniform carrier density is the most dominant broadening mechanism in our sample. 
We note that the inclusion of finite lifetime effects will also release the kinematic constraint on intra-band processes $s=s'$. For finite chemical potential it will yield a Drude-like contribution to $\chi''$ at low energy, $\omega \sim \Gamma_0$ \cite{Cardona}. We did not find any evidence for this contribution in our Raman data for $\omega>$80~cm$^{-1}$, but this contribution could be masked by the signal coming from the doped Si substrate.

\section{G band phonon data and ERS}

In the main text, the Dirac voltage is determined from the well-known gate-dependent evolution of the G-band energy and line-width. This determination can be cross-checked with the ERS continuum at 1580 cm$^{-1}$ which follows the evolution of the G band line-width. In figure \ref{fig3SI} we show the spectra of the G band, recorded at multiple gate voltages. They were performed on the same spot as the continuum data reported in the main text, but with a higher resolution using a 1800 grooves/mm grating. We fitted these spectra with Lorentzian profiles in order to extract the energy and line-width which are shown in figure \ref{fig3SI}(c). The Dirac voltage can be estimated from the minimum (maximum) of the phonon energy (line-width), yielding $V_{Dirac}\sim 20~V$. As shown in fig. \ref{fig3SI}(d), the evolution of the G band energy and line-width with Fermi energy can also be well reproduced theoretically by the well-known expressions for the G-band renormalization by interband electron-hole excitations \cite{Ando}
\begin{figure*}
 \includegraphics [width=0.85\textwidth] {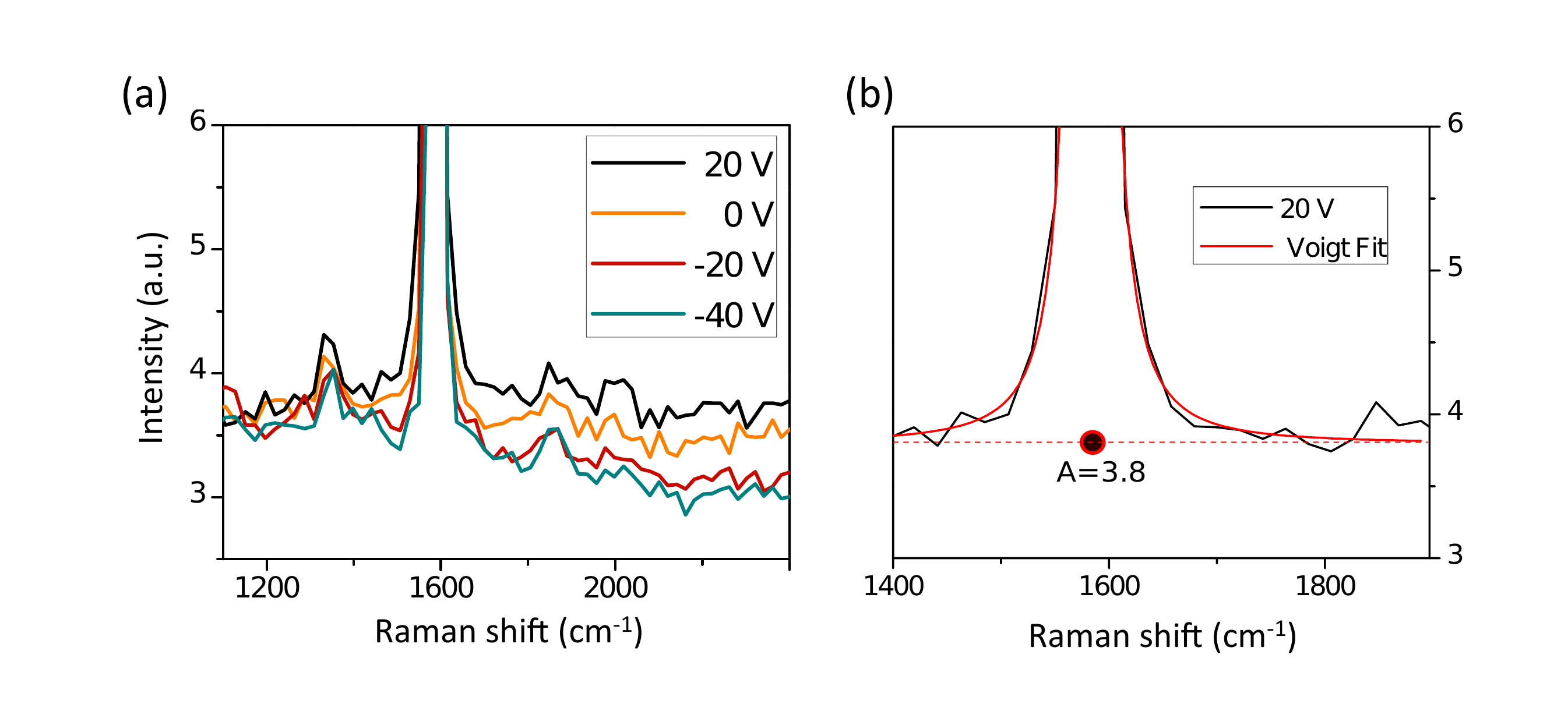}
 \caption{(a) Details of the ERS spectra near the G band energy recorded at different gate voltages. (b) Voigt fits with an additional constant background $A$ for the V$_{g}$=20~V spectrum}
 \label{fig4SI}
\end{figure*}
\par
 In figure 3 of the main text, the intensity of the ERS continuum at $\omega_G$ is plotted as a function of the gate voltage. It was determined using a lower resolution (600 groove/mm) in order to increase the photon counts on the continuum. This gave rise to a resolution limited line-shape for the G-band phonon (fig \ref{fig4SI}(a)). The G band was fitted with a Voigt profile together with an additional constant $A$ which describe the total continuum background (fig \ref{fig4SI}(b)) (for some gate voltages a additional linear in $\omega$ term was added to account for the continuum close to $\omega_G$). The constant $A$ is equal to the ERS intensity at $\omega_G$ up to a gate independent background $I_0$: $A$=I$_{ERS}$($\omega_G$)+$I_0$. $I_0$ was then identified as the value of $A$ far away from the Dirac voltage and subtracted in order to obtain the gate dependence of I$_{ERS}$($\omega_G$) shown on the main text (Fig. 3).

\section{Estimation of the 2E$_F$ position from the experimental Ratio R}

\begin{figure}
 \includegraphics [width=0.5\textwidth] {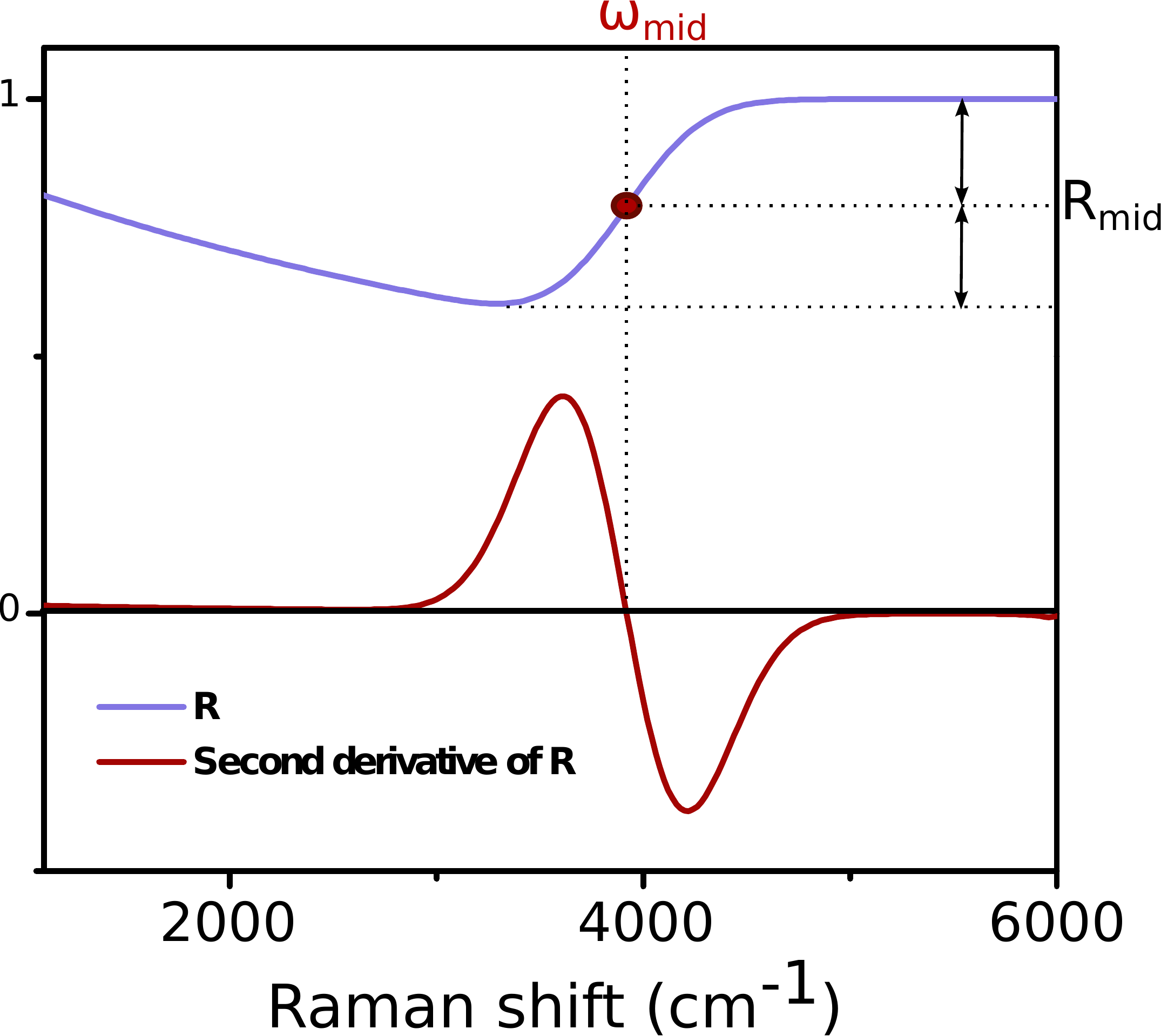}
 \caption{Determination of the 2E$_F$=$\omega_{mid}$ position as the inflexion point illustrated on the theoretical ratio $R (\omega, V_G)$}
 \label{fig6SI}
\end{figure}

As we can see in fig. 2(b) of the main text, after the introduction of the Fermi level broadening, the value of 2$E_F$ does not correspond to a clear onset of the suppression of the signal as in the clean homogeneous case at T=0K. Instead the onset becomes smeared out and 2E$_F$ now corresponds to the inflexion point of the curves (zero of the second derivative of either $I_{ERS}$ or $R$ as can be seen in Fig. \ref{fig1SI} and \ref{fig2SI}). While the second derivatives of the experimental spectra $R$ are relatively noisy, a more practical estimation of the position of the inflexion point can be performed from the $R (\omega, V_G)$ data by determining the energy $\omega_{mid}$, which corresponds to the mid-point $R_{mid}$ between the minimum of $R$ and $R$=1. The method is illustrated in figure \ref{fig6SI}. it was used for the data points of fig. 2(d) of the main text and also for the data on the second device (see below).  

\begin{figure*}
 \includegraphics [width=0.85\textwidth] {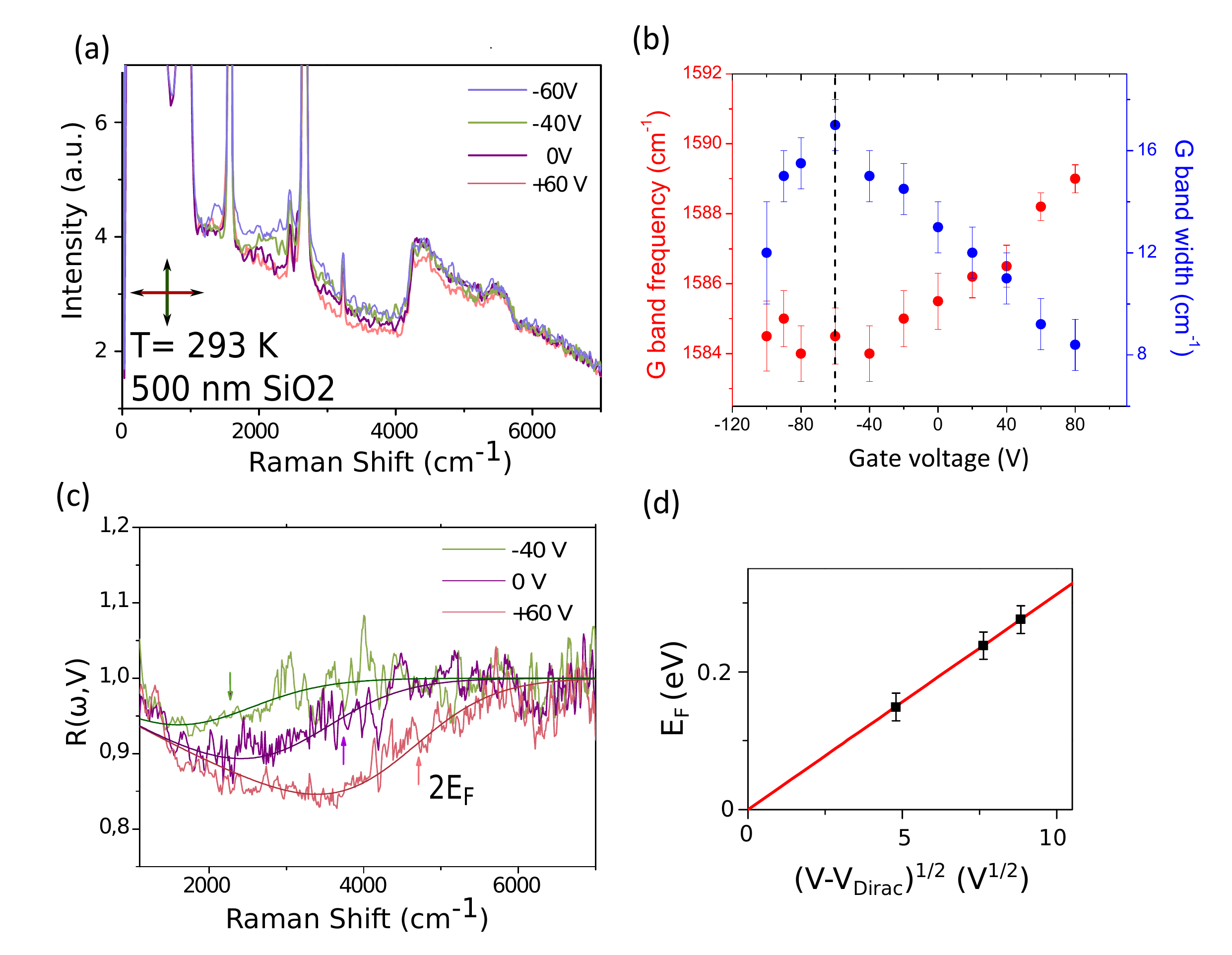}
 \caption{(a) Cross-polarization spectra recorded at four different gate voltages in the 0-7000~cm$^{-1}$ energy range at room temperature (T=293~K). (b) Gate dependance of the G band energy and line-width, allowing the determination of the Dirac voltage value ($V_{Dirac}$$\sim$ -60~V). (c) Experimental and theoretical gate dependence of $R(\omega, V_G)$ for $T = 293~ K$ and $C \sim 60~aF/\mu m^{2}$. A gaussian distribution of Fermi energy with $\delta E_F$=50~meV was used.
(d) Fermi energy plotted as a function of the square root of the gate voltage and experimental data (dots) extracted from the spectra in (c) by using the mid-point procedure described above to determine 2$E_F$ (marked by arrows in (a)).}
 \label{fig5SI}
\end{figure*}

\section{ERS data on a second graphene device}

In this paragraph we present additional ERS results on a second device. They are shown in figure \ref{fig5SI}. For this device the spectra were recorded at room temperature (T=293~K) in a vacuum better than 10$^{-5}$~mbar. The monolayer graphene sample was transferred on the top of a substrate with 500~nm of SiO$_2$ on doped Si, giving to the device a capacitance $C \sim 60~aF/\mu m^{2}$. The smaller capacitance due to the increased thickness of the dielectric was compensated by the fact that the insulating layer was more resistive, which allowed us to achieve higher gate voltage values.
Figure \ref{fig5SI} (a) shows cross-polarization spectra recorded at different gate voltages in the same experimental conditions (except the temperature) as the data in the first device presented in the main text. We can notice a background of similar overall intensity but with a slightly different shape with respect to the device shown in Fig 1(b) of the main text. Differences at high and low energy in particular seem to be associated with different substrate background for the two devices. We note that the overall shape of the background is also be controlled by interference conditions associated to SiO$_2$ thicknesses (280~nm for the first device versus 500~nm for the second), and by the optical alignment which could be slightly different between the two experiments.. Nonetheless the gate effect on the continuum is clearly observed in this device and, more importantly, the ratio $R$ shown in fig \ref{fig5SI} (c) can also be very well reproduced by the theoretical expectations using the estimated capacitance of the device $C$, T=293~K and the same distribution of Fermi energy as the first device.



\end{document}